\documentclass[aps,prx,superscriptaddress,reprint]{revtex4-2}
\usepackage{amsmath}
\usepackage{amssymb}
\usepackage{graphicx}
\usepackage{url}
\usepackage{hyperref}
\usepackage{color,xcolor}
\usepackage{epstopdf}
\usepackage{float}
\usepackage{ulem}
\usepackage[percent]{overpic}
\usepackage{siunitx}
\newcommand{\vect}[1]{\boldsymbol{#1}}

\begin{document}

\title{Magnetic phase diagram of a two-orbital model for \\ bilayer nickelates varying doping}
\author{Ling-Fang Lin}
\author{Yang Zhang}
\affiliation{Department of Physics and Astronomy, University of Tennessee, Knoxville, Tennessee 37996, USA}
\author{Nitin Kaushal}
\affiliation{Department of Physics and Astronomy and Quantum Matter Institute, University of British Columbia, Vancouver, British Columbia, Canada}
\author{Gonzalo Alvarez}
\affiliation{Computational Sciences and Engineering Division, Oak Ridge National Laboratory, Oak Ridge, Tennessee 37831, USA}
\author{Thomas A. Maier}
\affiliation{Computational Sciences and Engineering Division, Oak Ridge National Laboratory, Oak Ridge, Tennessee 37831, USA}
\author{Adriana Moreo}
\author{Elbio Dagotto}
\affiliation{Department of Physics and Astronomy, University of Tennessee, Knoxville, Tennessee 37996, USA}
\affiliation{Materials Science and Technology Division, Oak Ridge National Laboratory, Oak Ridge, Tennessee 37831, USA}

\begin{abstract}
Motivated by the recently discovered high-$T_c$ bilayer nickelate superconductor La$_3$Ni$_2$O$_7$, we comprehensively research a bilayer $2\times2\times2$ cluster for different electronic densities $n$ by using the Lanczos method. We also employ the random-phase approximation to quantify the first magnetic instability with increasing Hubbard coupling strength, also varying $n$. Based on the spin structure factor $S(q)$, we have obtained a rich magnetic phase diagram in the plane defined by $n$ and $U/W$, at fixed Hund coupling, where $U$ is the Hubbard strength and $W$ the bandwidth.
We have observed numerous states, such as A-AFM, Stripes, G-AFM, and C-AFM. For half-filling $n=2$ (two electrons per Ni site, corresponding to $N$ = 16 electrons), the canonical superexchange interaction leads to a robust G-AFM state $(\pi,\pi,\pi)$ with antiferromagnetic couplings in plane and between layers. By increasing or decreasing electronic densities, ferromagnetic tendencies emerge from the ``half-empty'' and ``half-full'' mechanisms, leading to many other interesting magnetic tendencies. In addition,  the spin-spin correlations become weaker both in the hole or electron doping regions compared with half-filling. At $n = 1.5$ (or $N=12$), density corresponding to La$_3$Ni$_2$O$_7$, we obtained the ``Stripe 2'' ground state (antiferromagnetic coupling in one in-plane direction, ferromagnetic coupling in the other, and antiferromagnetic coupling along the $z$-axis) in the $2\times2\times2$ cluster. In addition, we obtained a much stronger AFM coupling along the $z$-axis than the magnetic coupling in the $xy$ plane. The random-phase approximation calculations with varying $n$ give very similar results as Lanczos, even though both techniques are based on quite different procedures. Meanwhile, a state with $q/\pi = (0.6, 0.6, 1)$ close to the E-phase wavevector is found in our RPA calculations by slightly reducing the filling to $n=1.25$, possibly responsible for the E-phase SDW recently
observed in experiments. Our predictions can be tested by chemically doping  La$_3$Ni$_2$O$_7$.
\end{abstract}

\maketitle

\section{Introduction}
Since the discovery of superconductivity in Sr-doped infinite-layer NdNiO$_2$ films~\cite{Li:Nature}, nickelates became the newest member of the family of high-$T_c$ superconductors~\cite{Pan:nm,Botana:prx,Nomura:rpp,Zhang:prb20,Lee:prb04,Gu:innovation}, following the copper- and iron-based families~\cite{Bednorz:Cu,Dagotto:Rmp94,Kamihara:jacs,Dagotto:Rmp}. Very recently, a record $T_c$ $\sim 80$ K superconductivity was reported in bilayer La$_3$Ni$_2$O$_7$ under pressure \cite{Sun:arxiv}, opening a new remarkable avenue for the study of unconventional superconductivity~\cite{LiuZhe:arxiv,Zhang:arxiv-exp,Hou:arxiv,Yang:arxiv09,Zhang:arxiv09,Wang:arxiv9,Dong:arxiv12,Zhang:2402,Zhang:2333,Zhang:326,Wang:cplreview,Li:cpl24,Oh:prb23,Xue:cpl24,Takegami:prb,Jiang:prl,Sakakibara:arxiv09,Craco:prb24,Btzel:prb24,Kakoi:prb24,Chen:prb24,Geisler:2401.04258,Schl:arxiv23,Yang:arxiv23-dmrg,Lange:prb24,Kaneko:prb24,Huang:prb23,Btzel:prb24}.

\begin{figure}
\centering
\includegraphics[width=0.48\textwidth]{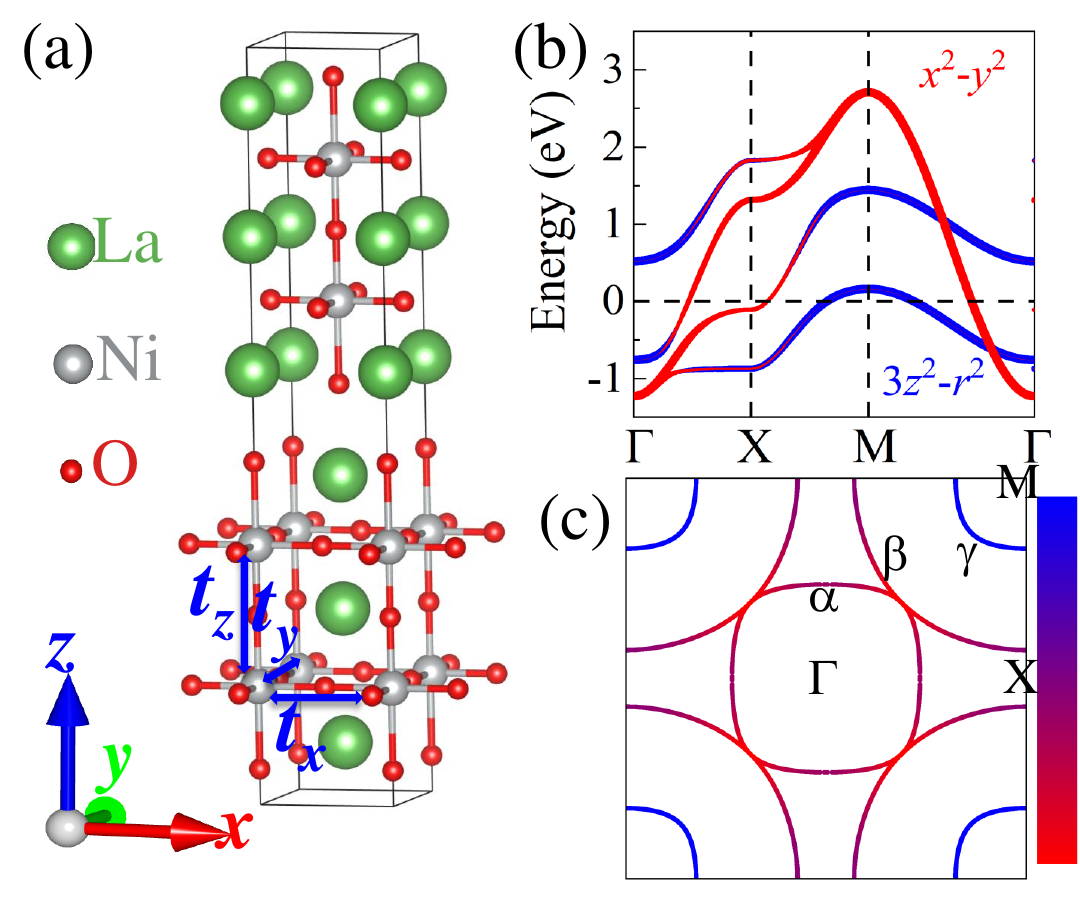}
\caption{(a) Crystal structure of La$_3$Ni$_2$O$_7$ in the Fmmm phase. The three directions of nearest-neighbor hopping matrices are indicated by $t_x$, $t_y$ and $t_z$. (b-c) Calculated (b) band structure and (c) Fermi surface with average electronic filling $n = 1.5$ per site for the two-orbital tight-binding model with the nearest-neighbor hopping along the $x-$, $y-$ and $z-$axis in the high pressure phase.}
\label{lattice_structure}
\end{figure}

La$_3$Ni$_2$O$_7$ has a bilayer NiO$_6$ octahedron stacking layered structure~\cite{Ling:jssc}, belonging to the Ruddlesden-Popper perovskite family, different from the NiO$_4$ layered nickelates\cite{Li:Nature,Pan:nm}. At ambient pressure, La$_3$Ni$_2$O$_7$ has an orthorhombic Amam (No. 63) atomic structure with an ($a^-$-$a^-$-$c^0$) out-of-phase octahedron tilting distortion around the [110] axis originating from the high-symmetry I4/mmm phase~\cite{Ling:jssc}. By introducing hydrostatic pressure, a first-order phase transition from Amam to Fmmm (No. 69) phases was obtained, fully suppressing the NiO$_6$ octahedron tilting rotations around 10 GPa (see Fig.~\ref{lattice_structure}(a)). In this symmetric phase, superconductivity was found in a broad pressure range from 14 to 43.5 GPa~\cite{Sun:arxiv}. In addition, the tetragonal I4/mmm phase has also been proposed to represent the high-pressure structure in both theory~\cite{Geisler:qm} and experiments~\cite{Wang:jacs}. However, those two phases provide basically the same physics~\cite{Sakakibara:prl24} because the distortion from I4/mmm to Fmmm is very small~\cite{Zhang:1313}.

The early report of superconductivity under pressure was based on a sharp drop and flat stage in resistance vs temperature, found by using KBr as the pressure-transmitting medium \cite{Sun:arxiv}. Moreover, a diamagnetic response in the susceptibility was reported, and those results
were interpreted as an indication of both zero resistance and Meissner effect~\cite{Sun:arxiv}, suggesting a superconducting state. Zero resistance was later confirmed by several studies~\cite{Hou:arxiv,Zhang:arxiv09,Zhang:arxiv-exp}. Very recently, the Meissner effect of the superconducting state was observed in the $ac$ magnetic susceptibility, with the superconducting volume fraction up to $48 \%$~\cite{Li:arxiv24}, suggesting bulk superconductivity, as opposed to initial suggestions of filamentary superconductivity \cite{Zhou:arxiv}.

First-principles density functional theory (DFT) calculations suggest that La$_3$Ni$_2$O$_7$ has a large charge-transfer gap between Ni's $3d$ and O's $2p$ orbitals and can be described as a Ni two-orbital bilayer model~\cite{Luo:prl23,Zhang:prb23}. Induced by canonical ``dimer'' physics~\cite{Zhang:prb23}, the Ni $d_{3z^2-r^2}$ orbital forms a bonding-antibonding molecular-orbital state, while the $d_{x^2-y^2}$ orbital remains decoupled between layers. Due to the in-plane hybridization between $e_g$ orbitals, the electronic occupation of both orbitals are nonintegers~\cite{Luo:prl23,Christiansson:prl23}. In addition, the Fermi surface of La$_3$Ni$_2$O$_7$ under pressure consists of two electron sheets ($\alpha$ and $\beta$), combinations of mixed $e_g$ orbitals, and a hole-pocket $\gamma$, made of the $d_{3z^2-r^2}$ orbital~\cite{Luo:prl23,Zhang:prb23}. This $\gamma$ pocket was not found in recent angle-resolved photoemission spectroscopy experiments for the non-superconducting Amam phase of La$_3$Ni$_2$O$_7$ ~\cite{Yang:arxiv09}. The random-phase approximation (RPA) many-body calculations suggest that this $\gamma$ hole pocket, and its associated Fermi surface nesting, plays the key role in mediating superconductivity in La$_3$Ni$_2$O$_7$~\cite{Zhang:nc24}. In addition, recent DFT calculations also found that the electron-phonon coupling alone is not sufficient to obtain such high $T_c$ in La$_3$Ni$_2$O$_7$ under pressure~\cite{Yi:arxiv2024,Ouyang:arxiv2024,Zhang:arxiv2024}, but superconductivity could be enhanced by combining the electron-phonon coupling and electronic correlations~\cite{Zhang:arxiv2024}.

Based on the two-orbital model described above, a $s_{\pm}$-wave pairing superconducting channel induced by spin fluctuations was reported by several theoretical studies, caused by the partial nesting of the Fermi surface in the high-pressure phase of La$_3$Ni$_2$O$_7$~\cite{Liao:prb23,Zhang:prb23-2,Zhang:nc24,Yang:prb23,Sakakibara:prl24,Gu:arxiv,Shen:cpl,Liu:prl23,Qu:prl,Lu:prl}. Although the $s_{\pm}$-wave pairing superconducting channel was considered to be driven by strong interlayer coupling, the $d_{x^2-y^2}$ orbital also has robust contributions to the superconducting gap, comparable to those of the $d_{3z^2-r^2}$ orbital~\cite{Zhang:nc24,Tian:prb24}. Alternative studies suggest a dominant $d_{x^2-y^2}$-wave superconducting pairing channel arising from the superconducting pairing state in the $\beta$ sheet~\cite{Lechermann:prb23,Jiang:cpl24,Fan:arxiv23}, driven mostly by the intralayer coupling, where this $\beta$ sheet is similar to that in the optimally doped cuprate superconductor. In addition, by introducing longer-range hoppings, the $d_{xy}$ symmetry pairing channel could also compete with the $s_{\pm}$-wave symmetry~\cite{Liu:arxiv2023,Heier:prb24}, also driven by the intralayer coupling. In fact, the interplay between the intralayer and the interlayer Cooper pairing was originally discussed by $t-J$ models in coupled plane or ladder geometries, finding that a strong interlayer coupling leads to 
$s$-wave pairing, while strong intralayer coupling would result in $d$-wave pairing~\cite{Dagotto:prb92}.

In addition, La$_3$Ni$_2$O$_7$ displays rich and complex density wave behaviors of spin and charge, depending on the pressure and temperature~\cite{Liu:scpma,Wu:prb01,Chen:RIXS,Xie:SB,Khasanov:arxiv}. Early experiments involving resistivity measurements reported a kink-like transition behavior around 153~K at ambient pressure, suggesting the existence of a spin-density wave (SDW)~\cite{Liu:scpma}. A recent muon-spin relaxation ($\mu$SR) experiment revealed the presence of SDW in La$_3$Ni$_2$O$_{7-\delta}$ without pressure~\cite{Khasanov:arxiv,Chen:prl24}. Furthermore, a nuclear magnetic resonance (NMR) experiment also supports the existence of SDW in La$_3$Ni$_2$O$_7$~\cite{Dan:arxiv24}. Neutron~\cite{Xie:SB} and resonant inelastic X-ray scattering~\cite{Chen:RIXS} experiments arrive at similar conclusions. However, due to the difficulty with high-pressure experiments, the full evolution of the magnetic properties of La$_3$Ni$_2$O$_7$ under pressure is still unclear.

In this work, using a Lanczos method, we systematically studied a bilayer $2\times2\times2$ cluster for several electronic densities $n$ \cite{footnote}. We obtained a very rich magnetic phase diagram varying $n$, involving the A-AFM, Stripe 2, G-AFM, and C-AFM phases. At half-filling with number of electrons 
$N = 16$ (or density $n=2$), due to the canonical superexchange antiferromagnetic (AFM) interaction, the system displays a robust G-AFM state with both AFM couplings in plane and between planes, in the bilayer geometry. By hole or electron doping away from half-filling, ferromagnetic (FM) tendencies start developing from the ``half-empty'' and ``half-full'' mechanisms \cite{Lin:prl21}. Thus, FM competes with AFM tendencies, leading to a rich magnetic phase diagram. Furthermore, the spin-spin correlations become weaker with increasing hole or electron doping away from half-filling. At $N = 12$ ($n=1.5$), corresponding to the electronic density of Ni in La$_3$Ni$_2$O$_7$, we obtained a Stripe 2 ground state (AFM in one in-plane direction and FM in the other, while AFM along the $z$-axis) in our $2\times2\times2$ cluster calculations, in agreement with multiple previous studies using other approximations. This agreement suggests that our predictions for other densities (reachable by doping La$_3$Ni$_2$O$_7$) are robust. In addition, we also obtained a much stronger AFM coupling along the $z$-axis than the magnetic coupling in the $xy$ plane. Complementing our Lanczos study, here we also employ the RPA varying the density $n$, reporting the leading magnetic instability with increasing Hubbard strength at several fixed densities. Overall, there is very good qualitative agreement between Lanczos and RPA, two very different approximations, again providing a robust foundation to our conclusions.

\section{Methods}

In this work, we employed the standard multi-orbital Hubbard model on a cube system ($2\times2\times2$), which includes a kinetic energy component and Coulomb interaction energy terms
$H = H_k + H_{int}$. The tight-binding (TB) kinetic portion is described as:
\begin{eqnarray}
H_k = \sum_{\substack{<i,j>\sigma\gamma\gamma'}}t_{\gamma\gamma'}(c^{\dagger}_{i\sigma\gamma}c^{\phantom\dagger}_{j\sigma\gamma'}+H.c.)+ \sum_{i\gamma\sigma} \Delta_{\gamma} n_{i\gamma\sigma},
\end{eqnarray}
where the first term represents the hopping of an electron from orbital $\gamma$ at site $i$ to orbital $\gamma'$ at site $j$. $\gamma$ and $\gamma'$ represent the two different orbitals. The second term includes the crystal-fields splitting terms.

The electronic interaction portion of the Hamiltonian is (using standard notation):
\begin{eqnarray}
H_{int}= U\sum_{i\gamma}n_{i\uparrow \gamma} n_{i\downarrow \gamma} +(U'-\frac{J_{\rm H}}{2})\sum_{\substack{i\\\gamma < \gamma'}} n_{i \gamma} n_{i\gamma'} \nonumber \\
-2J_{\rm H}  \sum_{\substack{i\\\gamma < \gamma'}} {{\bf S}_{i\gamma}}\cdot{{\bf S}_{i\gamma'}}+J_{\rm H}  \sum_{\substack{i\\\gamma < \gamma'}} (P^{\dagger}_{i\gamma} P^{\phantom\dagger}_{i\gamma'}+H.c.).
\end{eqnarray}
The first term is the intraorbital Hubbard repulsion. The second term is the electronic repulsion between electrons at different orbitals, where the standard relation $U'=U-2J_{\rm H}$ is assumed due to rotational invariance. The third term represents the Hund's coupling between electrons occupying the Ni's $3d$ orbitals. The fourth term is the pair hopping between different orbitals at the same site $i$, where $P_{i\gamma}$=$c_{i \downarrow \gamma} c_{i \uparrow \gamma}$.

To study the multi-orbital Hubbard model, including quantum fluctuations beyond mean-field approximations, the many-body technique that we employed was the Lanczos method, supplemented by the density matrix renormalization group (DMRG) method~\cite{white:prl,white:prb}. In practice, we used the \texttt{SCS\char`_Lanczos} \cite{nitin:web} and DMRG++ software packages~\cite{Alvarez:cpc}. For simplicity, for even electron numbers, the targeting up ($N_{\rm up}$) and down ($N_{\rm dn}$) electrons are equal, corresponding to $S_z$=0 sector. While for odd electron numbers, the targeting $N_{\rm up}$=$N_{\rm dn}+1$, corresponding to $S_z$=1/2 sector. Both open-boundary conditions (OBC) and periodic-boundary conditions (PBC) were considered along the $xy$ directions, while along the $z$ axis it was always OBC as in bilayers. Due to the special structure of our cluster, the only difference in plane between PBC and OBC is that we doubled the values of the $t_x$ and $t_y$ hoppings for the PBC case. 1000 Lanczos steps and convergence criterium below $10^{-10}$ were set during our Lanczos calculations. At several points in parameter space, DMRG calculations were also used to confirm our Lanczos results. We found that the results from Lanczos and DMRG agree very well. During the DMRG calculation, at least $1200$ states were kept and up to $21$ sweeps were performed.

In the tight-binding term, we used the basis \{$d_{3z^2-r^2}$,$d_{x^2-y^2}$\} , here referred to as $\gamma$ =  \{0, 1\}, respectively. For simplicity, we only considered the nearest-neighbor hopping matrices along the three directions, which are obtained from
our previous first-principles study for La$_3$Ni$_2$O$_7$ under pressure~\cite{Zhang:prb23}:
\begin{equation}
\begin{split}
t^{x}_{\gamma\gamma'} =
\begin{bmatrix}
  -0.115   &  0.240\\
  0.240    &  -0.492
\end{bmatrix},\\
\end{split}
\end{equation}
\begin{equation}
\begin{split}
t^{y}_{\gamma\gamma'} =
\begin{bmatrix}
  -0.115   &  -0.240\\
  -0.240    &  -0.492
\end{bmatrix},\\
\end{split}
\end{equation}
\begin{equation}
\begin{split}
t^{z}_{\gamma\gamma'} =
\begin{bmatrix}
  -0.644   &  0.000\\
 0.00    &  0.000
\end{bmatrix}.\\
\end{split}
\end{equation}

All the hopping matrix elements are given in eV units. $\Delta_{\gamma}$ is the crystal-field splitting of orbital $\gamma$. Specifically, $\Delta_{0} =0$, $\Delta_{1} = 0.398$ eV. The total kinetic energy
bandwidth $W$ is 3.936 eV.

\section{Results}

\subsection{TB results}

Before discussing our Lanczos results, the calculated TB electronic structures corresponding to La$_3$Ni$_2$O$_7$ under pressure are briefly reviewed here. The Fermi energy is obtained by integrating the density of states for all $\omega$ until a number of 1.5 electrons per Ni site is reached. Based on the obtained Fermi
energy, a $4001\times4001$ $k$-mesh was used to calculate the Fermi surface. The odd character of 4001 ensures that the $\Gamma$ point (0,0) is included in the grid used.

As shown in Fig.~\ref{lattice_structure}(b), the $d_{3z^2-r^2}$ orbital displays the expected bonding-antibonding splitting, where the bonding state band goes slightly above the Fermi level, resulting in a $\gamma$ hole-pocket around the M point (see Fig.~\ref{lattice_structure}(c)). In addition, the Fermi surface also contains two electron sheets ($\alpha$ and $\beta$) formed from a mixture of $d_{3z^2-r^2}$ and $d_{x^2-y^2}$ orbitals caused by the in-plane hybridization of $e_g$ orbitals. Those results are in agreement with previous theoretical studies~\cite{Luo:prl23,Zhang:prb23,Yang:prb23,Liu:prl23}.

\subsection{Phase diagrams}
\begin{figure}
\centering
\includegraphics[width=0.48\textwidth]{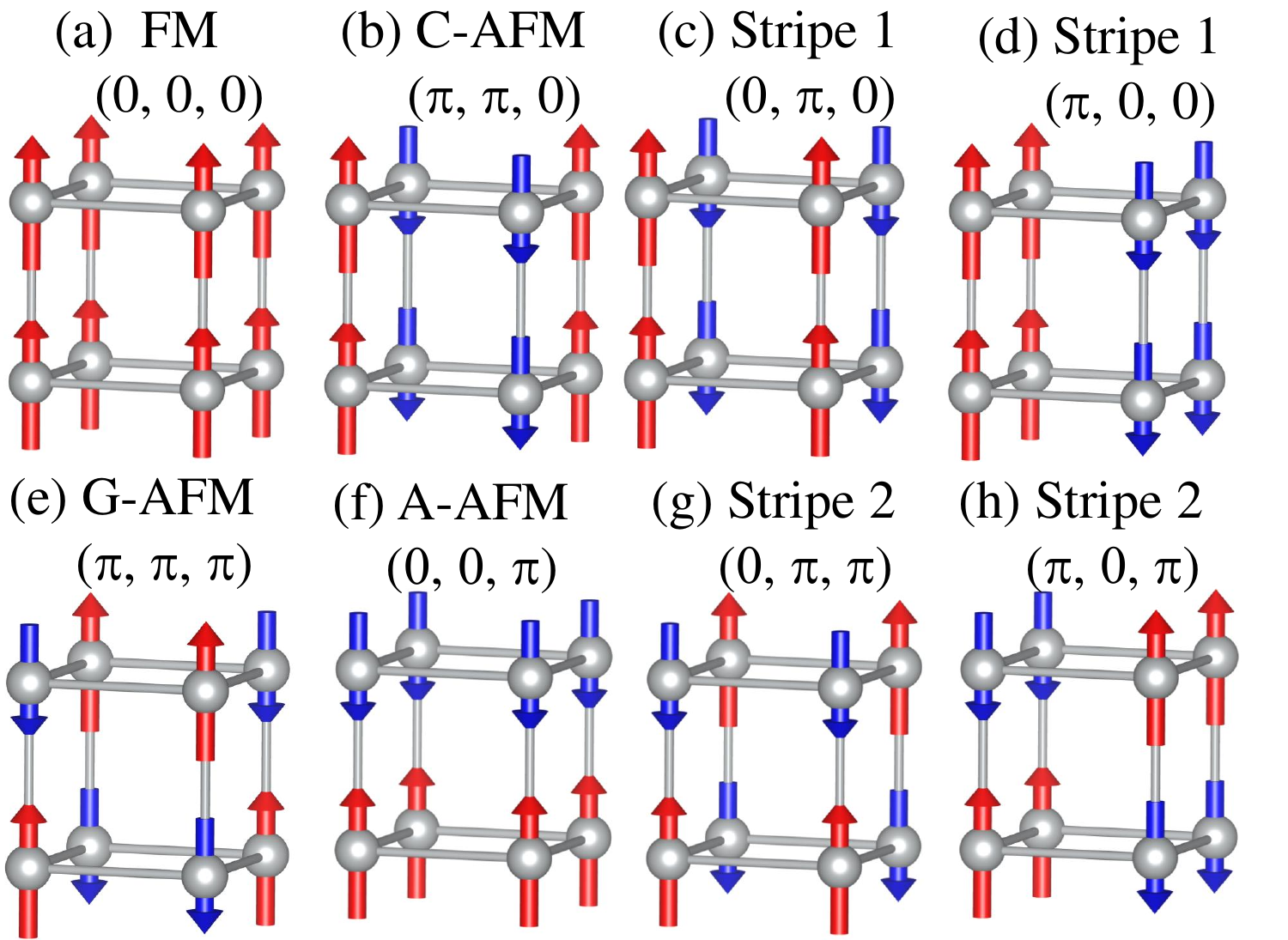}
\caption{Sketches of the observed magnetic structures for our $2\times2\times2$ system, including (a) FM, (b) C-AFM, (c) Stripe 1, (d) Stripe 1, (e) G-AFM, (f) A-AFM, (g) Stripe 2, and (h) Stripe 2, with the $S(\vect{q})$ peaks at $\vect{q}$ = (a) (0, 0, 0), (b) ($\pi$, $\pi$, 0), (c) (0, $\pi$, 0), (d) ($\pi$, 0, 0), (e) ($\pi$, $\pi$, $\pi$), (f) (0, 0, $\pi$), (g) (0, $\pi$, $\pi$), and (h) ($\pi$, 0, $\pi$), respectively.}
\label{magnetic_structure}
\end{figure}

To obtain the  magnetic phase diagram with different electronic densities and different 
electronic correlation strengths, using the Lanczos method we calculated the spin structure factor, which is defined as the Fourier transform of the spin-spin correlation function:
\begin{eqnarray}
S(\vect{q})=\frac{1}{N}\sum_{i,j}\langle{\vect{S}_i \cdot \vect{S}_j}\rangle e^{i\vect{q} \cdot (\vect{r}_i-\vect{r}_j)},
\end{eqnarray}
where $\vect{S}_i$  is the spin at site $i$, $\vect{r}_i$ is the position of site $i$, $\langle{\vect{S}_i \cdot \vect{S}_j}\rangle$ is the average of real space spin-spin correlation between sites $i$ and $j$, $N$ is the total number of lattice sites, and $\vect{q}$ is the wave vector.

\begin{figure*}
\centering
\includegraphics[width=0.98\textwidth]{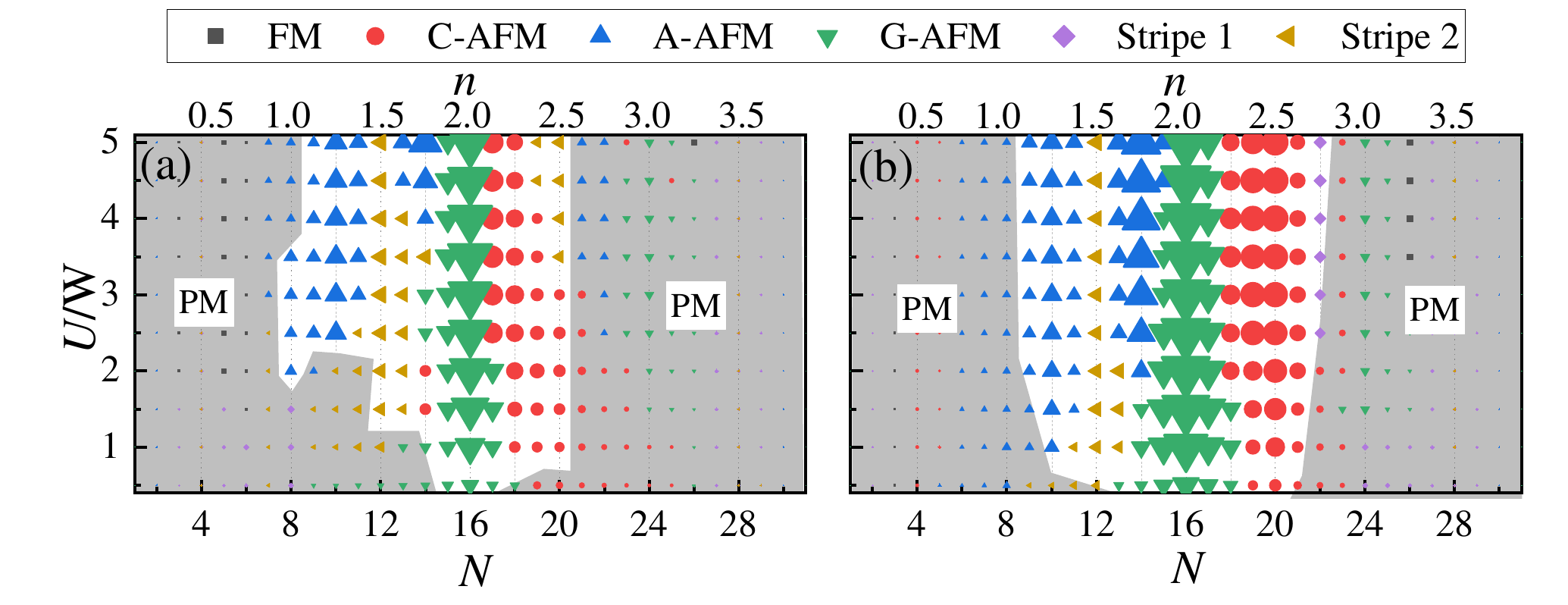}
\caption{Magnetic phase diagram of the small cluster studied here, varying $U$/W and $N$, at fixed $J_{\rm H}/U=0.2$. Different magnetic phases, including FM, C-AFM, A-AFM, G-AFM, Stripe 1, and Stripe 2, are indicated by different points with different shapes and colors. The size of the points is proportional to the strength of the spin structure factor $S(\vect{q})$. For simplicity, the regions with $S(\vect{q})\lesssim2$ are marked as PM by using a gray color [using the number 2 as cutoff appears arbitrary, but for comparison the largest $S(\vect{q})$ is $\approx 10$ at $U/W=5$, $N=16$, and OBC]. (a) is using PBC while (b) is using OBC.}
\label{phase_diagram_0.2}
\end{figure*}

We show all the magnetic structures observed for a $2\times2\times2$ system in Fig.~\ref{magnetic_structure}, which includes FM, C-AFM, Stripe 1 (two energy degenerate variations), G-AFM, A-AFM, Stripe 2 (two energy degenerate variations), corresponding to the $S(\vect{q})$ peak at $\vect{q}$= (0, 0, 0), ($\pi$, $\pi$, 0), (0, $\pi$, 0), ($\pi$, 0, 0), ($\pi$, $\pi$, $\pi$), (0, 0, $\pi$), (0, $\pi$, $\pi$), and ($\pi$, 0, $\pi$), respectively. Due to rotational symmetry, $S(0, \pi,  \pi)$ [$S(0, \pi,0)$] and $S(\pi, 0, \pi)$ [$S(\pi, 0,0)$] should be the same, which represents the Stripe 2 [1] phase.

Based on the measurements of the spin structure factor $S(\vect{q})$, we obtained the magnetic phase diagram for different electron numbers $N$ and electronic correlations $U/W$, at fixed $J_{\rm H}/U=0.2$ as in our previous studies \cite{Zhang:prb23}. Results are shown in Fig.~\ref{phase_diagram_0.2}. Here, we considered the cases PBC and OBC. The results are similar for the two boundary conditions. Several interesting magnetic states emerge in the phase diagrams. At the half-filling situation ($N = 16$), the G-AFM is dominant in a robust manner for all values of $U/W$ studied. By hole or electron doping away from the half-filling $N = 16$ case, the magnetic phases display a rich behavior due to the multi-orbital nature of the model. The states reached by doping involve paramagnetism (PM),  A-AFM, Stripe 2, C-AFM, and Stripe 1. In addition, the spin-spin correlations also become weaker in the hole or electronic doping regions, compared to the half-filling $N = 16$ case. Note that there is a slight difference between the cases of PBC and OBC. This is because in PBC, the $t_x/t_y$ needs to be considered doubled in strength, as already explained, but only once in OBC, leading to a stronger AFM vs. FM competition in the $xy$ plane in the PBC case.

\begin{figure*}
\centering
\includegraphics[width=0.98\textwidth]{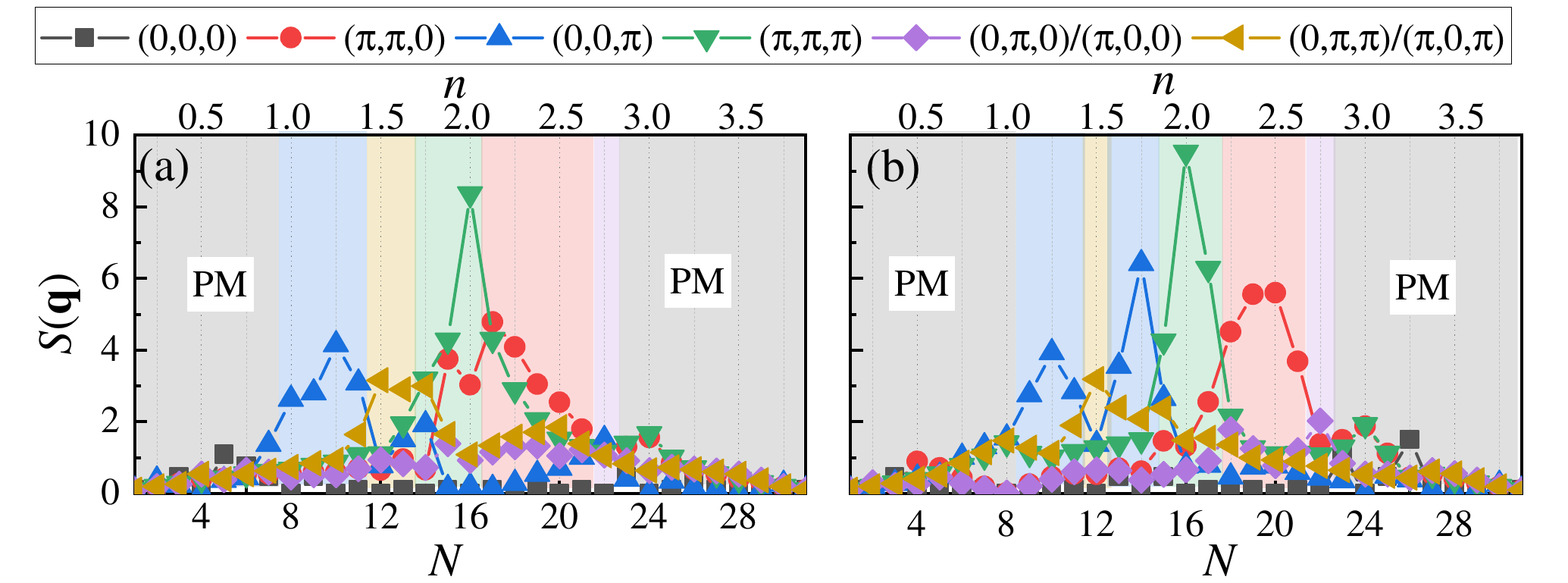}
\caption{The dominant spin structure factors for different spin states, as a function of different values of $N$ at $U/W =3$  for the cases of (a) PBC and (b) OBC.}
\label{UbyW3}
\end{figure*}

For the benefit of the readers, in Fig.~\ref{UbyW3} we also show the detailed spin structure factors $S(\vect{q})$ for different spin states vs $N$, all at $U/W$=3. For $N\lesssim8$ and $\gtrsim24$, the $S(\vect{q})$ is quite weak without substantial differences in peak strengths among different spin states, overall indicating PM. For other electronic densities, such $N =10, 12, 16$ and 18, the peaks of $S(\vect{q})$ are located at $\vect{q}=(0,0,\pi)$, $(0,\pi,\pi)/(\pi,0,\pi)$, $(\pi,\pi,\pi)$, and $(\pi,\pi,0)$, corresponding to the A-AFM, Stripe 2, G-AFM, and C-AFM phases, respectively. By reproducing the same analysis of $U/W = 3$ for other values of $U/W$, we obtained the phase diagrams in Fig.~\ref{phase_diagram_0.2}.

\begin{figure}
\centering
\includegraphics[width=0.48\textwidth]{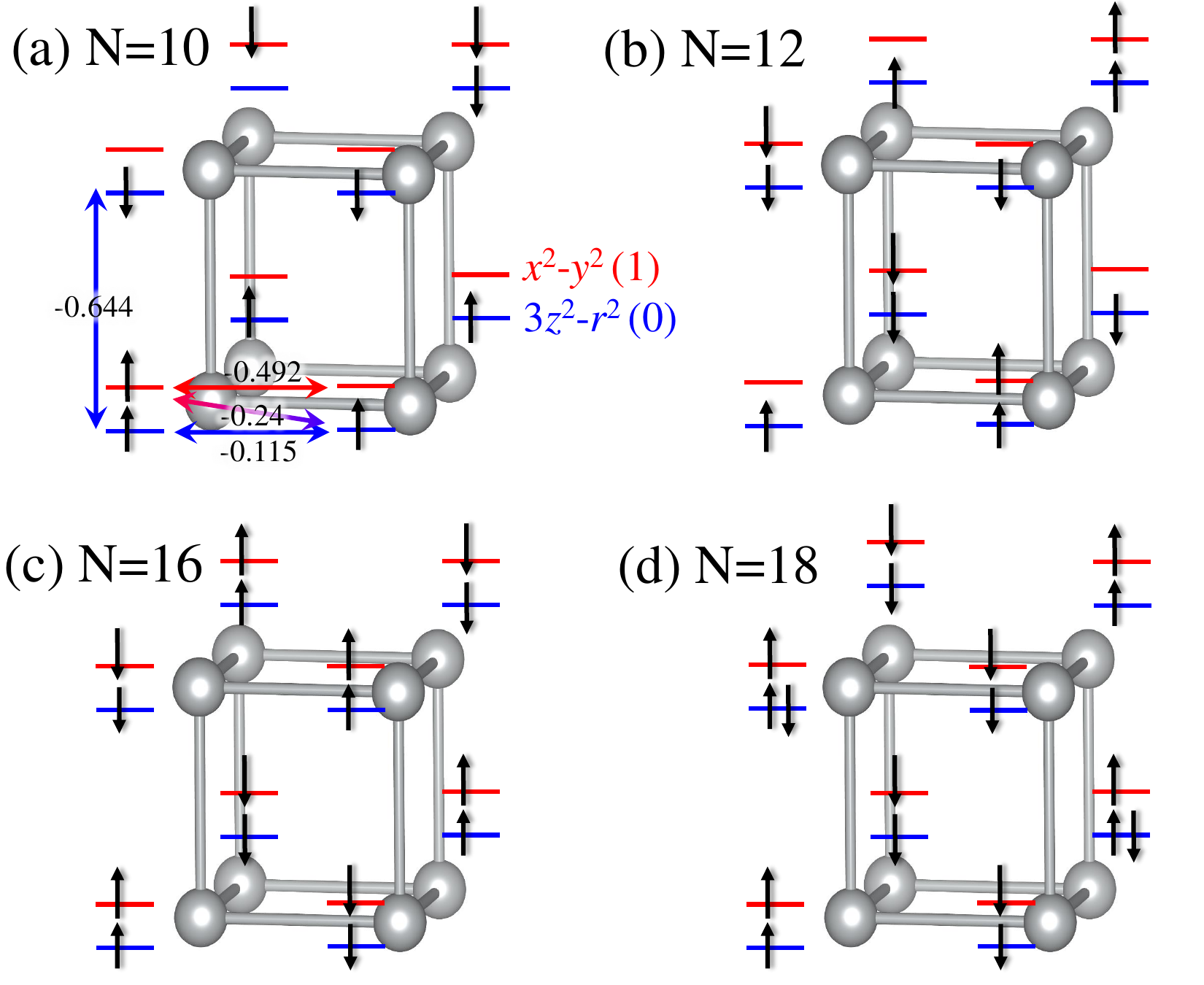}
\caption{Electronic configurations for some typical values of $N$, including (a) 10, (b) 12, (c) 16, and (d) 18. Here, the spin up and down are marked by arrows and the different orbitals are  distinguished by different colors.}
\label{intuitive_explanation}
\end{figure}

\subsection{Intuitive explanation of the phase diagram}

Let us discuss now the intuitive explanation for the existence of the main phases at several key number of electrons $N$=10, 12, 16 and 18. This can be understood by three main coupling mechanisms~\cite{Lin:prl21,Lin:cp,Lin:prb22}, based on two-site second-order perturbation theory: (1) half-empty FM coupling, (2) half-half AFM coupling, and (3) half-full FM coupling. Here, we also display the sketches of intuitive explanations of these main phases in Fig.~\ref{intuitive_explanation}. First, let us focus on the $z$-direction, for $N=10$, 12 and 16. Only the ``half-half'' AFM coupling mechanism (standard superexchange) is responsible for those cases as displayed in Figs.~\ref{intuitive_explanation}(a-c), thus the AFM coupling is obtained, corresponding to the $S(\vect{q})$ peak at $\pi$ along the $z$ direction. 

By increasing $N$, because of having more than two electrons per two orbitals per site, the ``half-full'' FM coupling mechanism starts influencing on the energy of the many phases, leading to the competition between FM and AFM bonds at various values of $N$, leading to the C-AFM phase at $N = 18$ (see, as example, the detailed second-order perturbation theory analysis for this case in Appendix section A).

Now let us analyze the situation for the $xy$-plane. For the half-filling $N = 16$ case, the canonical superexchange half-half AFM coupling leads to a robust G-AFM order. Correspondingly, the $S(\vect{q})$ peak is at ($\pi$, $\pi$) along the $xy$ direction. (1) In the hole doping region away from half-filling, the AFM coupling caused by superexchange strongly competes with the FM coupling caused by the ``half-empty'' mechanism \cite{Lin:prl21}, leading to an in-plane stripe phase (AFM in one direction and FM in the other), such as Stripe 2 at $N = 12$. At $N = 10$, the FM tendency wins over AFM tendency in plane, leading to in-plane FM coupling (AFM along $z$-axis as analyzed in the previous paragraph), resulting in a global A-AFM phase. (2) In the electron doping region away from half-filling,  due to the ``half-half'' mechanism \cite{Anderson:pr} AFM coupling wins, resuting at $N = 18$ in a C-AFM state (AFM coupling along the $xy$-plane and FM coupling along the $z$-axis).  Increasing $N$ further, the FM coupling caused by ``half-full'' mechanism starts playing an important role, competing with AFM coupling from ``half-half'' mechanism, resulting in the Stripe 1 phase (AFM in-plane in one direction and FM in the other, while it is FM along the $z$-axis) for OBC, and A-AFM and C-AFM phases for PBC at $N = 22$ (see Figs.~\ref{phase_diagram_0.2}(a) and (b)).

For the $N = 12$ case -- with average electronic density $n = 1.5$ per site corresponding to La$_3$Ni$_2$O$_7$ in our $2\times2\times2$ cluster calculations  -- we obtained a Stripe 2 phase [with wavevectors ($\pi$, 0, $\pi$) or (0, $\pi$, $\pi$)], in agreement with previous theoretical studies using RPA \cite{Zhang:prb23-2,Zhang:nc24}. This is quite remarkable, namely studying a small system and solving it exactly, gives very similar conclusions as RPA for the realistic electronic density $n=1.5$, with RPA work done by us and other groups. Note that RPA is based on a totally different many-body technique, sum of an infinite subset of Feynman diagrams as compared to solving exactly a small cluster.


Furthermore, we found that the AFM spin-spin correlations along the $z$-axis are much stronger than the spin-spin correlations along the $xy$ plane, which is in good agreement with recent experiments \cite{Chen:RIXS,Xie:SB}. For example, for the $N = 12$ case with $U/W = 3$, $J_{\rm H}/U=0.2$, $\langle{\vect{S}_i \cdot \vect{S}_j}\rangle = -0.897$ along $z$-axis while it is $-0.054$ along $x$- and $y$-direction. The reasons are (1) the hoppings along $z$ direction are larger than the ones in $xy$ plane, (2) due to the degeneracy of the ($\pi$, 0, $\pi$) and (0, $\pi$, $\pi$) states, the $xy$ plane nearest spin-spin correlations have up-down, down-up, up-up and down-down cases,  so they cancelled out with each other.

It should be noticed that for the real samples of La$_3$Ni$_2$O$_7$, experiments found oxygen deficiencies \cite{Dong:arxiv12}, leading to ``effectively'' varying the electronic density in the Ni site, away from $n=1.5$. Thus, the AFM coupling in between planes  should be very robust against oxygen deficiency. But along the in-plane directions, we found a strong FM vs. AFM competition even just slightly varying the number of electrons (such as Stripe 2 at $N = 12$ and A-AFM at $N = 13$), suggesting that the oxygen deficiency may seriously affect the magnetic coupling along the in-plane directions in real samples. 

Note that for a larger size system, unreachable using Lanczos due to the exponential increase in the size of the Hilbert space, the magnetic phase diagram could be even richer due to in-plane FM and AFM competition. Specifically, due to these size limitations in clusters that can be studied with Lanczos, we could not explore other complex magnetic states, such as the E-AFM \cite{Sergienko:prl}, CE-AFM \cite{Yunoki:prl} and block states \cite{Zhang:prb045119}. In fact, very recent DFT calculations suggested that the E-AFM ground state dominates at ambient pressure~\cite{LaBollita:E-AFM,Zhang:E-AFM,Ni:E-AFM}. Such a more complex structure as the E-AFM phase cannot fit into our small cluster, thus we cannot confirm or deny the existence of this state and other. Alternative (all approximate) many-body techniques,
like unrestricted Hartree-Fock \cite{Lin:prm,Daghofer:prb,Pandey:prb})  will be needed to address this issue. The E-phase notation, widely used in manganites, is often referred to as double-stripe in nickelates.

In summary, in spite of their limitations, our calculations clearly reveal (1) Stripe 2 dominance at $n=1.5$ as found via other many-body techniques, (2) a much stronger AFM coupling along the $z$-axis than along the $xy$ plane, (3) strong AFM and FM competition in the $xy$ plane when moving slightly away from $n=1.5$, (4) reduced spin spin correlations as compared with half-filing $N = 16$, and (5) a rich phase diagram varying $N$ that merits experimental confirmation via chemical doping.

\subsection{RPA Results}
We have also used the RPA method to quantify the magnetic properties of this bilayer system, varying the electronic density $n$. Past analyses have shown that the RPA, which is based on a perturbative weak-coupling expansion in the Hubbard interaction \cite{Kubo2007,Graser2009,Altmeyer2016,Romer2020}, captures the essence of the physics for Fe-based and Cu-based superconductors. Reference~\cite{Graser2009} provides background information, and we expect the RPA then to succeed also for Ni-based superconductors.

Specifically, we analyze the RPA enhanced spin susceptibility tensor $\chi({\bf q}, \omega=0)$ that is obtained from the Lindhart function tensor $\chi_0({\bf q})$ as
\begin{eqnarray}
\label{eqn}
\chi({\bf q}) = \chi_0({\bf q})[1-{\cal U}\chi_0({\bf q})]^{-1}.
\end{eqnarray}
Here, $\chi_0({\bf q})$ is an orbital-dependent susceptibility tensor and ${\cal U}$ is a tensor containing the interaction parameters \cite{Graser2009}.

\begin{figure}
\centering
\includegraphics[width=0.48\textwidth]{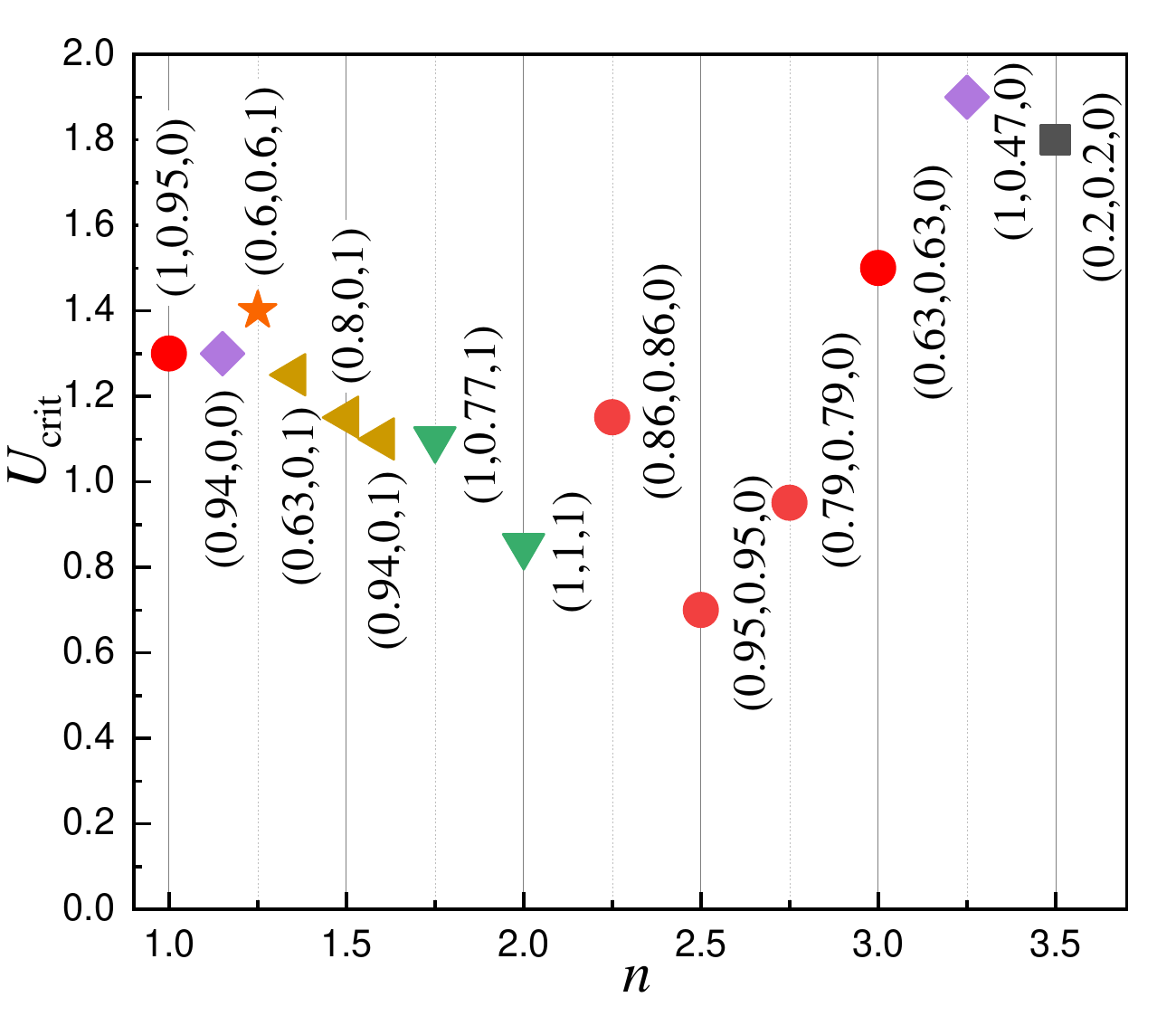}
\caption{RPA leading states with their vector $q$ (in vertical parenthesis, in units of $\pi$) and $U_{\rm crit}$ vs. $N$, where $U_{\rm crit}$ is the critical $U$ for the magnetic instability within RPA.}
\label{RPA}
\end{figure}

Figure~\ref{RPA} shows $U_{crit}$, the critical value of $U$, for which the susceptibility $\chi({\bf q})$ in Eq.~\ref{eqn} diverges, as a function of filling $n$, and the corresponding wavevector $q$ of the instability. We note that, unlike in the Lanzcos, in the RPA this SDW instability generally occurs for incommensurate wavevector ${\bf q}$. At $n=1.5$, the RPA found a state  with ${\bf q} = (0.8, 0, 1)$ in units of $\pi$, thus resembling the Stripe 2 state of the Lanczos results with ${\bf q} = (1, 0, 1)$ in units of $\pi$. At larger $n$, the RPA finds states close to the G-AFM, C-AFM, and FM states as $N$ increases, in close agreement with our Lanczos phase diagram. In the RPA, we do not find the A-AFM state for smaller $n$. However, as shown in Fig.~\ref{phase_diagram_0.2}(a), in the Lanzcos this state primarily occurs for $U$/W $\gtrsim 2$, i.e. much larger $U$ than the $U_{\rm crit}$ of the RPA and beyond the range of $U$ where the weak coupling RPA is applicable.  In fact, for $U \lesssim U_{\rm crit}$, where the RPA is expected to give accurate results, there is excellent agreement between both techniques. Furthermore, we also found a state with ${\bf q} = (0.6, 0.6, 1)$ close to the E-phase wavevector by slightly reducing the filling to $n=1.25$, which is possibly responsible for the SDW discussed in recent experiments \cite{Chen:RIXS,Dan:arxiv24}.

\section{Conclusions}

In summary, motivated by the recently discovered bilayer nickelate superconductor La$_3$Ni$_2$O$_7$, here we theoretically studied a bilayer $2\times2\times2$ cluster by using the Lanczos method, as well as the RPA in the weak-coupling region. A remarkably rich magnetic phase diagram was obtained, involving a variety of phases, such as the A-AFM, Stripe 2, Stripe 1, G-AFM, and C-AFM phases. Overall, this unusual variety is caused by the strong competition between FM and AFM tendencies at different electronic densities $n$ per Ni. For $n = 1.5$, corresponding to La$_3$Ni$_2$O$_7$, in our $2\times2\times2$ cluster calculations we obtained the Stripe 2 ground state (AFM in one in-plane direction, FM in the other, while AFM along the $z$-axis). Furthermore, we also obtained a much stronger AFM coupling along the $z$-axis than the magnetic coupling in the $xy$ plane. In addition, we observed a strong AFM vs FM competition in the $xy$ plane. Note that if larger cluster sizes were available for Lanczos, other even more complex interesting phases could become stable, such as the E-AFM, CE-AFM, and block states, as extensively discussed in the case of manganites and iron-based superconductors~\cite{Sergienko:prl,Yunoki:prl,Zhang:prb045119}. Meanwhile, a state with 
${\bf q} = (0.6, 0.6, 1)$ close to the E-phase wavevector  is found in our RPA calculations by slightly reducing the filling to $n=1.25$, which is possibly responsible to the SDW as discussed in the recent experiments. By hole or electronic doping away from half-filling $n = 2$, the spin-spin correlations become weaker as the doping increases in both directions. Thus, our results provide clear predictions of novel states for future experiments on Ni-oxide bilayer superconductors once the electronic density is varied via proper chemical doping.

\textit{Note added}. After completing our work, we noticed that a very recent independent theoretical work also showed similar results by using the RPA method varying the density $n$~\cite{Zhang:arxiv24}.

\section{Acknowledgments}
We appreciate inspiring discussions with B. Pandey and P. Laurell. This work was supported by the U.S. Department of Energy, Office of Science, Basic Energy Sciences, Materials Sciences and Engineering Division. G. A. was supported by the U.S. Department of Energy, Office of Science, National Quantum Information Science Research Centers, and Quantum Science Center. G. A. contributed his expertise with the DMRG algorithm, its applicability to the multi-orbital $2\times2\times2$ cube, and the software implementation.

\section{Appendix}

\subsection{Second-order perturbation theory analysis for $N = 18$}

\begin{figure}
\centering
\includegraphics[width=0.4\textwidth]{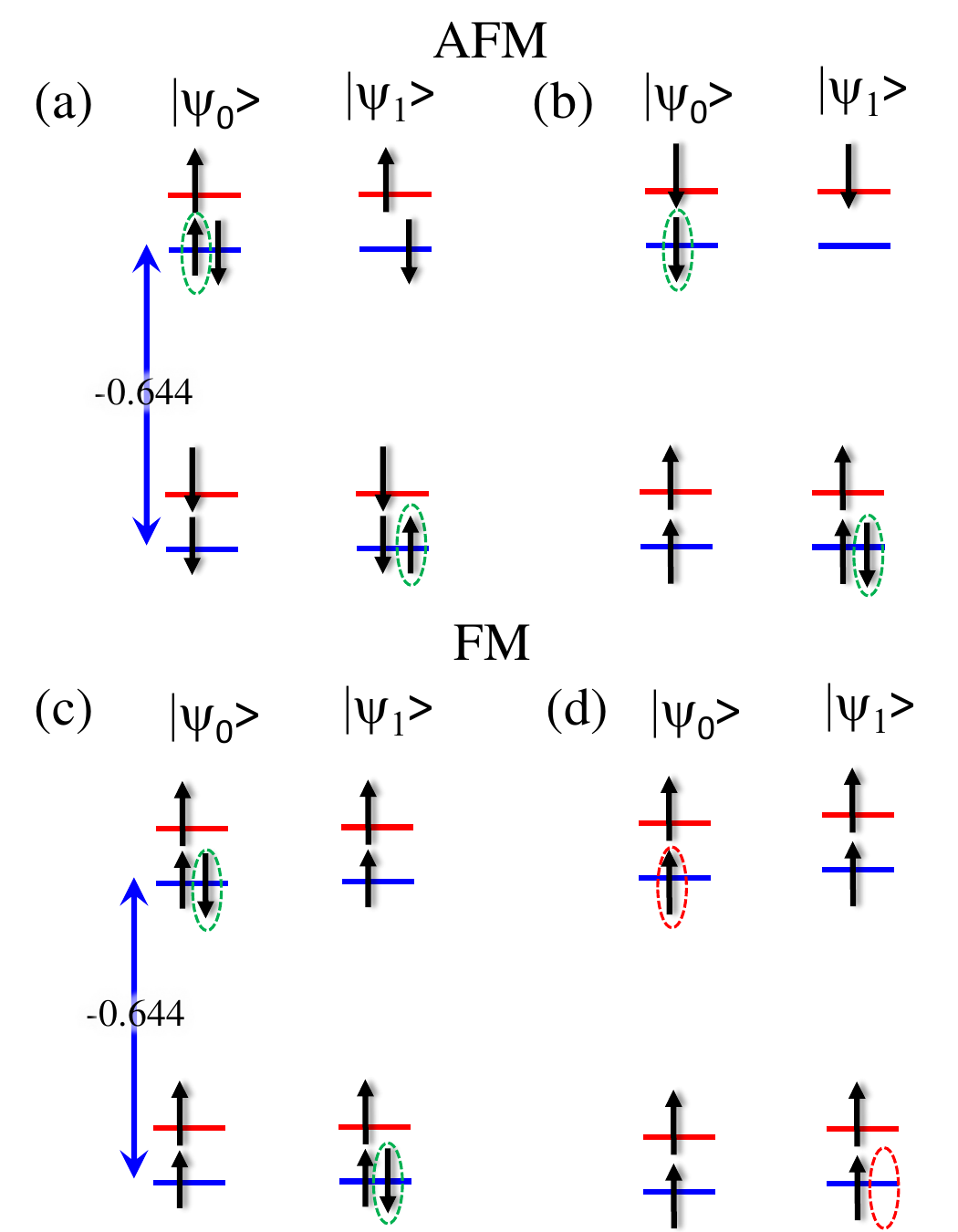}
\caption{Sketches of the virtual hopping process for $N$=18 along the $z$ direction. For simplicity, 2 sites with 2 orbitals are considered for both the AFM and FM cases. (a) and (c) are for 5 electrons, while (b) and (d) are for 4 electrons. Note that for the case (d), the hopping is forbidden (red dash ovals) due to the Pauli exclusion principle.}
\label{n18z}
\end{figure}

For the benefit of the readers, we include here our study for $N$=18, with focus on the $z$ direction, to show the detailed two-site second-order perturbation theory analysis in the atomic limit, as illustrated in Fig.~\ref{n18z}. Here the hopping $t^{z}_{00}=-0.644$ eV is treated as perturbation in the large $U/W$ limit and we focus on the effect of the $U$, $U'$, and Hund's coupling terms.
For non-degenerate states (cases (a) and (b)), it is well known that second-order perturbation theory always lowers the energy by:
\begin{equation}
\Delta E=-\sum_{\substack{m}}\frac{|\langle \psi_m|H'|\psi_0\rangle |^2}{E_m-E_0}.
\end{equation}

For case (a), the ground state and first excited state atomic energies, starting point of the perturbation theory, are
\begin{equation}
E_0=U+3U'-J_{\rm H},
\end{equation}

\begin{equation}
E_1=U+3U'+J_{\rm H},
\end{equation}
respectively.

By using second-order perturbation theory, it can be shown that the energy gain of the AFM configuration (case (a)) due to $ t^{z}_{00}$ is
\begin{equation}
\begin{split}
\Delta E_{\rm (a)}&=\frac{|\langle \psi_1|H'|\psi_0\rangle |^2}{E_0-E_1}\\
&=-\frac{|t^{z}_{00}|^2}{2J_{\rm H}}.
\end{split}
\end{equation}

Repeating the calculation for case (b), the ground state and excited state atomic energies are
\begin{equation}
E_0=2U'-2J_{\rm H},
\end{equation}

\begin{equation}
E_1=U+2U',
\end{equation}
respectively.

Then, the energy gain of the AFM configuration for case (b) is
\begin{equation}
\begin{split}
\Delta E_{\rm (b)}&=\frac{|\langle \psi_1|H'|\psi_0\rangle |^2}{E_0-E_1}\\
&=-\frac{|t^{z}_{00}|^2}{U+2J_{\rm H}}.
\end{split}
\end{equation}

For the FM case (c), the two energy eigenstates of the unperturbed Hamiltonian are degenerate
\begin{equation}
E_0=E_1=U+3U'-J_{\rm H}.
\end{equation}
Thus, we should mix those degenerate eigenstates to obtain the energy difference as
$$
\begin{bmatrix}
E_0 & t^{z}_{00} \\
t^{z}_{00} & E_0 \\
\end{bmatrix}
\begin{bmatrix}
\psi_0 \\
\psi_1\\
\end{bmatrix}
=E\begin{bmatrix}
\psi_0 \\
\psi_1\\
\end{bmatrix}.
$$

We obtain $E_{\pm}=E_0\pm t^{z}_{00}$, so $\Delta E_{\rm (c)}$= $t^{z}_{00}$.

For the FM case (d), hopping is forbidden due to the Pauli principle, so there is no energy gain.

By using $U/W =3$ and the hoppings in the Method section, at $J_{\rm H}/U =0.2$, the total energy gained in the AFM state from $t^{z}_{00}$ is $\sim 0.113$ eV, which is smaller than the total energy gained from the FM state which is 
$\sim 0.644$ eV. This result indicates the FM case wins over AFM. Thus, along the $z$-axis a transition from wavevector $\pi$ for hole doping to wavevector $0$ for electron doping is expected, in excellent agreement with the Lanczos results.

\subsection{Lanczos phase diagrams at $J_H/U = 0.1$}

\begin{figure*}
\centering
\includegraphics[width=0.98\textwidth]{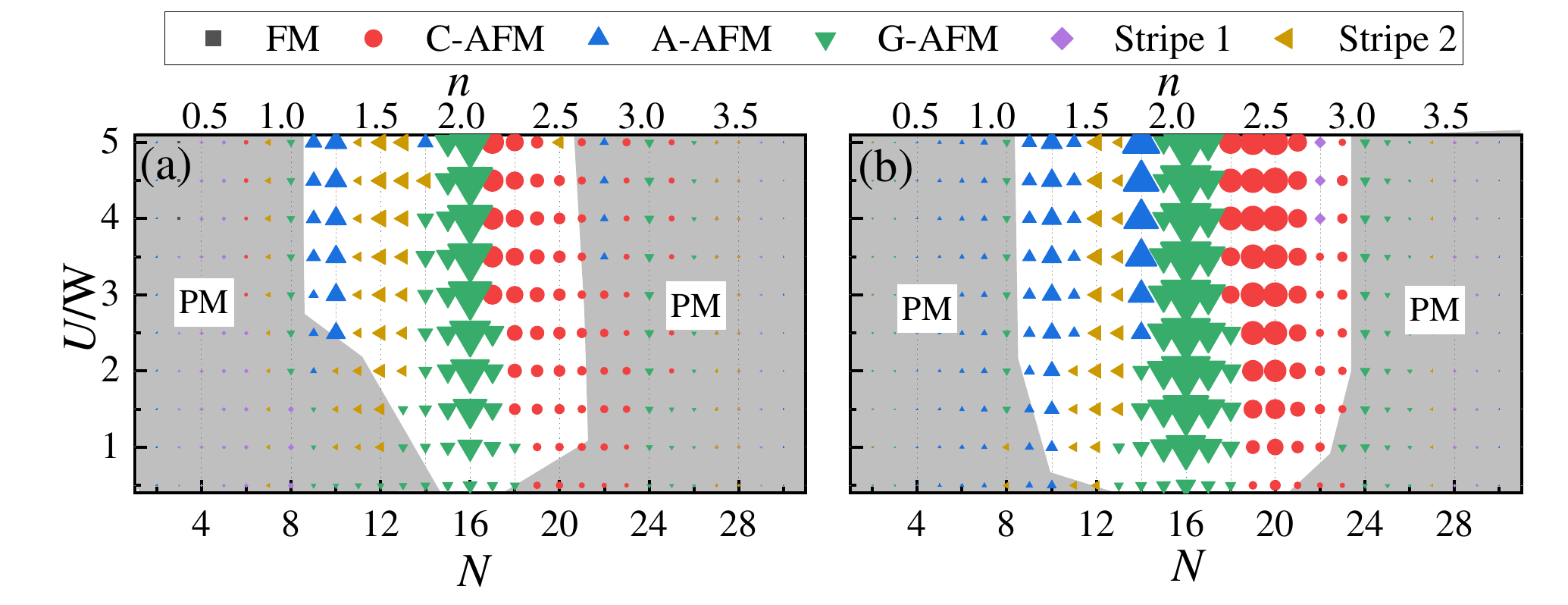}
\caption{Magnetic phase diagram of the small cluster studied here, varying $U$/W and $N$, at fixed $J_{\rm H}/U=0.1$. Different magnetic phases, including FM, C-AFM, A-AFM, G-AFM, Stripe 1, and Stripe 2, are indicated by different points with different shapes and colors. The size of the points is proportional to the strength of the spin structure factor $S(\vect{q})$. For simplicity, the regions with $S(\vect{q})\lesssim2$ are marked as PM by using a gray color. (a) is using PBC while (b) is using OBC.}
\label{phase_diagram_0.1}
\end{figure*}

As shown in Fig.~\ref{phase_diagram_0.1}, we also calculated the Lanczos phase diagrams at $J_{\rm H}/U = 0.1$. Those results were found to be similar to the case $J_{\rm H}/U=0.2$, indicating that our overall results are robust against changes in $J_{\rm H}/U$.

\subsection{Lanczos phase diagrams at n = 1.5} 
\begin{figure*}
\centering
\includegraphics[width=0.98\textwidth]{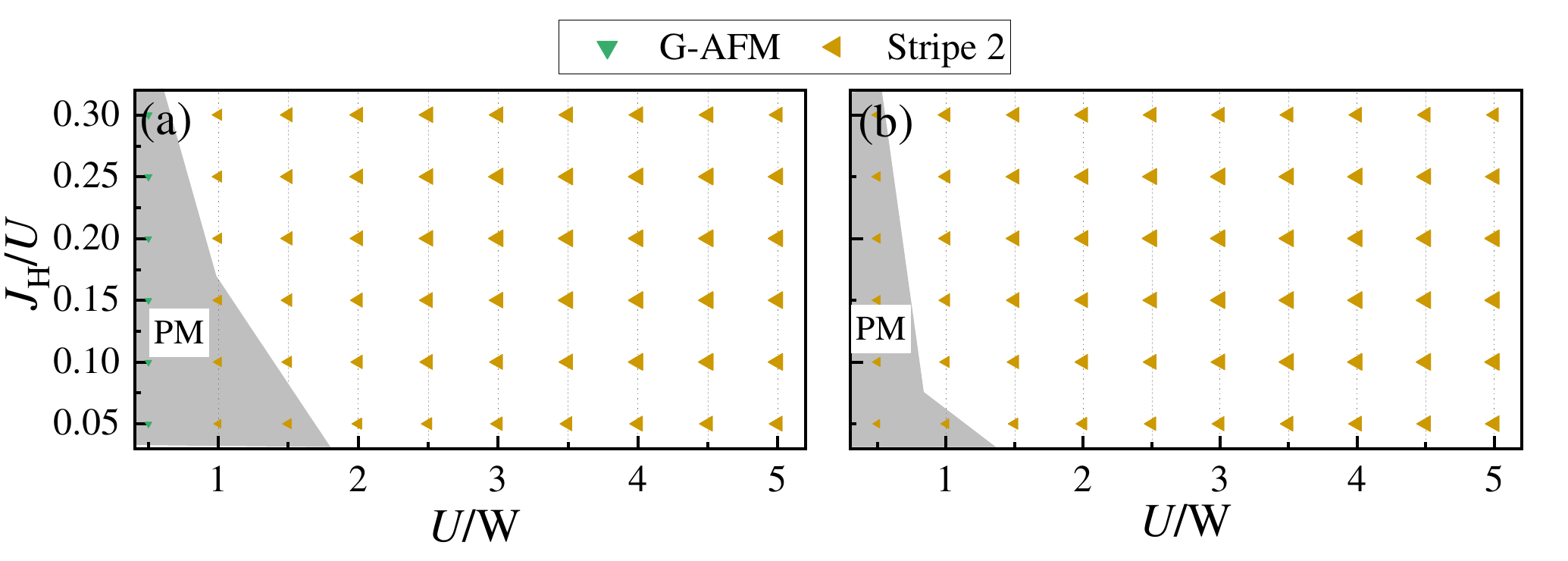}
\caption{Magnetic phase diagram of the small cluster studied here, varying $U$/W and $J_{\rm H}/U$ at fixed $N=12$. Different magnetic phases, including G-AFM, and Stripe 2, are indicated by different points with different shapes and colors. The size of the points is proportional to the strength of the spin structure factor $S(\vect{q})$. For simplicity, the regions with $S(\vect{q})\lesssim2$ are marked as PM by using a gray color. (a) is using PBC while (b) is using OBC.}
\label{phase_diagram_N12}
\end{figure*}

As shown in Fig.~\ref{phase_diagram_N12}, we also calculated the Lanczos phase diagrams by varying  $U$/W and $J_{\rm H}/U$ at fixed $N=12$. The Stripe 2 state is robust against by changing in $J_H/U$ and $U$/W.


\begin{references}
\bibitem{footnote} We have verified that exact diagonalization of the half-filled one orbital Hubbard model in a $2\times2$ cluster correctly indicates the $(\pi,\pi)$ AFM order obtained in larger clusters.
\end{references}
\end{document}